\documentclass[aps,twocolumn,pra,showpacs,10pt]{revtex4-1}

\usepackage{epsfig,amssymb,amsmath,graphicx}

\begin{document}


\title{Imaging of quantum Hall states in ultracold atomic gases}
\author{James S. Douglas and Keith Burnett}
\affiliation{University of Sheffield, Western Bank, Sheffield S10 2TN, United Kingdom}
\begin{abstract}
We examine off-resonant light scattering from ultracold atoms in the  quantum Hall regime. When the light scattering is spin dependent, we show that images formed in the far field can be used to distinguish states of the system. The spatial dependence of the far-field images is determined by the two-particle spin-correlation functions, which the images are related to by a transformation.
Quasiholes in the system appear in images of the density formed by collecting the scattered light with a microscope, where the quasihole statistics are revealed by the reduction in density at the quasihole position.
\end{abstract}
\pacs{67.85.-d, 73.43.-f, 42.50.Ct, 42.30.Va}
\maketitle


Electrons that are confined to two dimensions in a magnetic field and cooled to within a few degrees of absolute zero behave in ways that have fascinated physicists for decades. In 1980 von Klitzing \textit{et al.}~performed a groundbreaking experiment with a cold electron system and observed plateaus in the Hall resistance as the magnetic field was varied \cite{Klitzing1980a}.
Remarkably, the Hall resistance was exactly quantized at these plateaus to a value of $h/(ne^2)$, where $n$ is a positive integer. This integer quantum Hall effect is understood in terms of noninteracting electrons and is due to complete filling of the single-particle electron orbitals known as Landau levels. 

This discovery was soon followed by the observation of a similar effect, where the Hall resistance was exactly quantized as $h/(\nu e^2)$, with $\nu$ a fraction \cite{Tsui1982a}. This fractional quantum Hall effect occurs when the Landau levels are partially filled, with filling factor $\nu$, and the interactions between the electrons drive the gas into a strongly correlated state. This effect cannot be understood using perturbation theory and much of the current understanding is due to proposal of trial wave functions, such as the celebrated Laughlin wavefunction \cite{Laughlin1983a}, given by
\begin{equation}
\Psi_n = \prod_{j<l}^{N}(z_j-z_l)^n e^{-\sum_u|z_u|^2/2}.
\label{eq:Laughlin_state}
\end{equation}

The integer and fractional quantum Hall effects were originally observed in electron systems where the electron spin was polarized as a result of the Zeeman energy associated with the magnetic field. However, Halperin found that in some cases the Zeeman energy would be smaller than the other energy scales involved and that low-energy states with reversed spins could exist at some filling factors \cite{Halperin1983a}. Halperin proposed a spin-singlet state with half the electrons in each spin state as a potential ground state of the system. Halperin's intuition was confirmed by numerical studies of small electron systems that showed at filling factor $\nu = 2/5$ for vanishing Zeeman energy the exact ground state is unpolarized and has good overlap with the Halperin state \cite{Chakraborty1984a,Rezayi1987a}. 
 
Currently experimental studies of quantum Hall states use semiconductor materials that have their properties defined and limited by their composition. This makes it difficult to study the quantum Hall effect as system parameters are varied. A more flexible realization of these effects may be possible in ultracold, neutral atomic gases \cite{Bloch2008a, Lewenstein2007a}. Synthetic magnetic fields can be induced for these neutral gases by rotation \cite{Cooper2008a} or by optical means \cite{Lin2009a, Dalibard2010a}, potentially providing access to the quantum Hall regime. It is hoped that the purity of systems of ultracold atoms and the ability to dynamically control system parameters will enable new insights into the nature of the quantum Hall effect and allow novel states to be achieved, such as those with non-Abelian excitations that may be used for topological quantum computing \cite{Nayak2008a}.

The usefulness of simulating the quantum Hall effect in ultracold gases will be limited by our ability to gain information about states of the system, and effective means of probing the system must be developed. A number of methods have been suggested. Characteristics of the wave function may be revealed by time-of-flight imaging \cite{Read2003a,Lukin2005a,Palmer2008a} or density profile measurement \cite{Cooper2005a}. Excitation of edge states reveals information about the systems topological order \cite{Cazalilla2005a,Stanescu2010a} and Bragg scattering 
has been proposed to further probe the excitations of the system
\cite{Hafezi2007a,Stanescu2010a,Liu2010b}. Density measurement in Fermi gases could be used as a probe of the effective conductivity \cite{Shao2008a,Liu2010a,Oktel2008a} and the statistics of quasiparticles might be probed in a Ramsey-type experiment \cite{Paredes2001a}. Here we examine what imaging based on light scattering reveals about an ultracold  gas in the quantum Hall regime while the atoms remain \textit{in situ}.


\section{Light scattering from multi-component states}

We consider a specific example of light scattering from a Bose gas, where the $N$ atoms are divided equally into two $m_F$ spin states of a particular hyperfine level, which we refer to as spin up and spin down.
In neutral atomic gases, synthetic magnetic fields can be created that only couple to the motional state of the atom and do not couple to the spin of the atom, unlike an actual magnetic field. This extra experimental degree of freedom could allow the quantum Hall regime to be realized with zero Zeeman effect. For a two component Bose gas, with an atomic interaction that does not depend on the internal state,  a candidate ground state is the bosonic version of the Halperin state \cite{Paredes2002a}, given by $\Psi_H = S[\psi]$, where
\begin{multline}
\psi = \prod_{k<l}^{N/2}(z_k-z_l)\uparrow_k\uparrow_l\prod_{N/2<r<s}^{N}(z_{r}-z_{s})\downarrow_r\downarrow_s\\
\times\prod_{i<j}^{N}(z_i-z_j) e^{-\sum_u|z_u|^2/2}
\label{sqhs_ground_state}
\end{multline}
and $S$ symmetrizes over the particle labels.
This state is expressed in terms of $z_j=(x_j + i y_j)/\ell$, where $\mathbf{x}_j = (x_j,y_j)$ are the coordinates of the atoms, and $\uparrow_j$, $\downarrow_j$ are the spin-up and spin-down spinors.  The characteristic length scale $\ell = \sqrt{\hbar/(e B)}$  is determined by the effective magnetic field $B$ generated by the optical field or rotation and is of the order of a few microns for typical parameters  \cite{Klein2005a}.  This bosonic Halperin state has associated fractional filling factor $\nu_1 = 2/3$.

When the atomic interaction becomes spin dependent, such that the interaction between like spins is much stronger than the interaction between opposite spins, a paired state can exist \cite{Paredes2002a}. Atoms of opposite spin pair together to form spin-triplets, resulting in a Pfaffian state 
\begin{equation}
\Psi_P = \rm{Pf}\left(\frac{\uparrow_i\uparrow_j+\downarrow_i\downarrow_j}{z_i-z_j}\right)\prod_{k<l}^{N}(z_k-z_l) e^{-\sum_u|z_u|^2/2}\label{eq:pfaffian_state},
\end{equation}
where the Pfaffian is defined for an $N\times N$ antisymmetric matrix $M$ by \cite{Moore1991a}
\begin{equation}
\rm{Pf}(M_{ij}) = \frac{1}{2^{N/2}(N/2)!}\sum_{\sigma\in \rm{S}_N}\rm{sgn}(\sigma)\prod_{k=1}^{N/2} M_{\sigma(2k-1),\sigma(2k)}.
\end{equation}
This Pfaffian state is closely related to the ground state for the BCS theory of electrons with spin-triplet $p$-wave pairing and represents a superconducting state of effective fermions \cite{Paredes2002a}.

Scattering of light has been suggested theoretically and used experimentally in a number of contexts to probe the spin distribution of ultracold atoms \cite{Carusotto2004a,Higbie2005a,Carusotto2006a,Eckert2007a,Vega2008a,Eckert2008a,Zhang2009a,Roscilde2009a,Javanainen2003,Ruostekoski2008a,Douglas2010a,Weitenberg2011a}. Here we wish to determine if light scattering is a useful probe of the many-body states described above. We consider experiments done with alkali-metal atoms where the detuning from the $S_{1/2} \rightarrow P_{1/2},P_{3/2}$ transitions is larger than the natural line widths and also larger than
the Rabi frequency of any particular radiation mode. In this case the evolution of the atomic excited states can be adiabatically eliminated and in the rotating wave and dipole approximations the light-matter interaction is described by \begin{equation}
\hat{H}
=\sum_{m_1,m_2}\int\! d\mathbf{r}\hat{\Psi}^\dagger_{m_2}(\mathbf{r})
(\mathbf{\tilde{E}}^-(\mathbf{r})\tensor{\alpha}_{m_1,m_2}\mathbf{\tilde{E}}^+(\mathbf{r}))\hat{\Psi}_{m_1}(\mathbf{r}),
\end{equation}
where $\tensor{\alpha}_{m_1,m_2}$ is the polarizability tensor  \cite{Hammerer2010a,Deutsch1998a}. Here the atomic field operators $\hat{\Psi}^\dagger_{m}(\mathbf{r})$ and $\hat{\Psi}_{m}(\mathbf{r})$ create and destroy atoms in the ground hyperfine state $|F m\rangle$, and $\mathbf{\tilde{E}}^+(\mathbf{r})$ and $\mathbf{\tilde{E}}^-(\mathbf{r})$ are the slowly varying positive- and negative-frequency components of the electric field.

The polarizability tensor can be decomposed into three parts \cite{Hammerer2010a}, which correspond to a spin-independent interaction that does not play a role in our imaging following the filtering procedure discussed below and two spin-dependent interactions. The second spin-dependent term goes to zero as the detuning is increased and is typically at least an order of magnitude smaller than the first and we may neglect it. We are then left with the following effective interaction
\cite{Vega2008a, Douglas2010a}
\begin{equation}
\hat{H}
=-a(\Delta)\int\! d\mathbf{r}\boldsymbol{\hat{\rho}}(\mathbf{r})\cdot(\mathbf{\tilde{E}}^-(\mathbf{r})\times\mathbf{\tilde{E}}^+(\mathbf{r})).
\label{eq:spin_interaction}
\end{equation}
where  $\boldsymbol{\hat{\rho}}(\mathbf{r})=\sum_{m_1,m_2} \hat{\Psi}^\dagger_{m_2}(\mathbf{r})\langle F m_2|\hat{\mathbf{F}}/\hbar|F m_1\rangle\hat{\Psi}_{m_1}(\mathbf{r})$ depends on the spin density of the atoms and $a(\Delta)$ is the detuning dependent coupling (see Hammerer \textit{et al.}~\cite{Hammerer2010a}). We assume in our example calculation that the two atomic spin states have $m_F=\pm 1$ and that the spin quantization axis is $\hat{\mathbf{z}}$.


\begin{figure}
\centering
\includegraphics{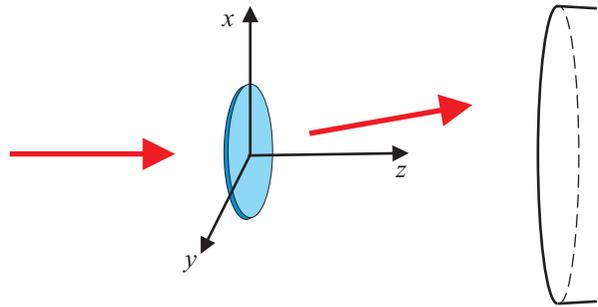}
\caption{(Color online) Light propagating in the $z$ direction is incident on the two-dimensional gas confined to the $xy$ plane. Light scatters and is collected by an imaging system on the $z$ axis.}
\label{light_scatter}
\end{figure}


We consider a two-dimensional atomic gas in the $xy$ plane that is illuminated by coherent light propagating in the $z$ direction, wave vector $k_0 \hat{\mathbf{z}}$, with linear polarization in the $y$ direction as shown in Fig.~\ref{light_scatter}. The interaction described by Eq.~(\ref{eq:spin_interaction}) then leads to scattering of light to wave vectors $\mathbf{k} = k(\sin\theta\cos\phi,\cos\theta,\sin\theta\sin\phi)$, where $\theta$ is the polar angle with respect to the $y$ axis and $\phi$ is the azimuthal angle in the $xz$ plane. The off-resonant nature of the scattering results in frequency shifts for the light of order of the atomic recoil frequency, which is many orders of magnitude less than the carrier frequency, and to a very good approximation we have $k = k_0$. The wave vector of the scattered light it then completely determined by $\boldsymbol\kappa$, the projection of $\mathbf{k}$ into the $xy$ plane. 
For light scattered predominately in the $z$ direction the interaction results in a rotation of the polarization, where we describe the scattered light in terms of polarization vectors
$\epsilon_1(\boldsymbol\kappa) = (\sin\phi, 0 ,-\cos\phi)$ and
$\epsilon_2(\boldsymbol\kappa) = (-\cos\theta\cos\phi,\sin\theta,-\cos\theta\sin\phi)$.
Placing a polarization filter with extinction axis in the $y$ direction on the $z$ axis will then remove the input light from the signal and will also filter the majority of light scattered into the mode with polarization $\boldsymbol\epsilon_2(\boldsymbol\kappa)$, as well as light scattered by the spin independent interaction \cite{Vega2008a}.

Image formation then depends on the rate of scattering to $\boldsymbol\kappa$ with polarization $\boldsymbol\epsilon_1(\boldsymbol\kappa)$. To first order in the coupling constant \cite{Douglas2011a}, this is proportional to $I(\boldsymbol\kappa) = \langle\varphi|\varphi\rangle$ where  
\begin{equation}
|\varphi\rangle = \int d\mathbf{x}\boldsymbol{\hat{\rho}}(\mathbf{x})
\cdot(\boldsymbol\epsilon_1(\boldsymbol\kappa)\times\hat{\mathbf{y}})e^{-i\boldsymbol\kappa\cdot\mathbf{x}}|\Psi\rangle
\end{equation}
for the many-body state $|\Psi\rangle$ and the integration has been reduced to two dimensions by taking into account the tight confinement in the $z$ direction. The first-order treatment of light scattering is expected to reproduce the same images as an experiment, provided the experimental images are averaged over a number of runs \cite{Douglas2011b}.
For scattered light propagating predominately in the $z$ direction, $\boldsymbol\epsilon_1(\boldsymbol\kappa)\times\hat{\mathbf{y}} = \hat{\mathbf{z}}\sin\phi+\hat{\mathbf{x}}\cos\phi \sim \hat{\mathbf{z}}$ and we can neglect the small corrections resulting from the $x$ component \cite{Douglas2010a}. The dominant features of an image then result from the $z$ component of the spin operator, $\hat{\rho}_z(\mathbf{x}) \equiv \sum_{m} m\int dz\hat{\Psi}^\dagger_{m}(\mathbf{r})\hat{\Psi}_{m}(\mathbf{r})$.


\begin{figure}
\centering
\includegraphics{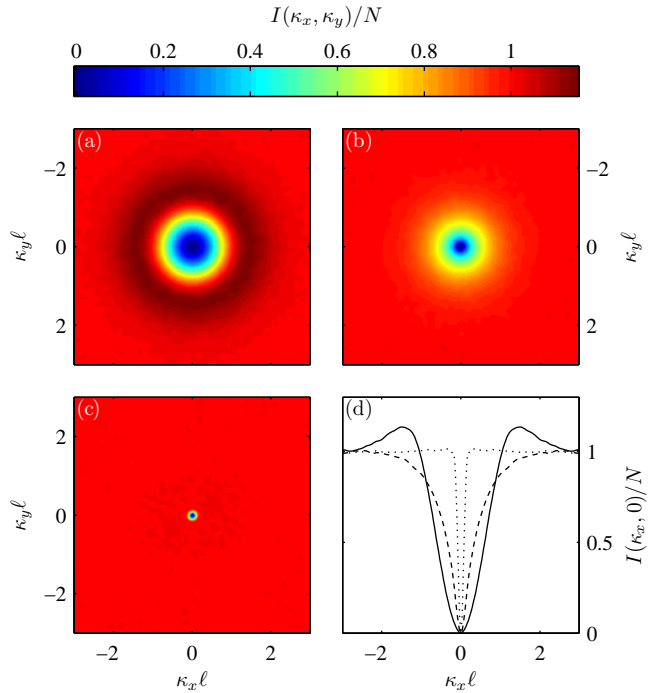}
\caption{(Color online) Far-field images of (a) the bosonic Halperin state, (b) the Pfaffian state, and (c) the comparison Laughlin state. (d) Slices through the far-field images at $\kappa_y=0$ for the bosonic Halperin state (solid), the Pfaffian state (dashed), and the comparison Laughlin state (dotted). Images are generated from 100000 Monte Carlo configurations of 100 particles.}
\label{sqs_far_field}
\end{figure}


One of the simplest methods of imaging the system is to let the scattered light propagate into the far field, where the distance from the sample is much greater than the wavelength of the imaging light. In this region the spatial dependence of the light intensity is simply determined by the Fourier components of the scattered light,  that is 
$I(\alpha \mathbf{x}) \propto  I(\boldsymbol\kappa)$, with scaling constant $\alpha$ that depends on the distance from the sample to the detector and determines the spatial size of the image. For a many-body state $|\Psi\rangle = \frac{1}{\mathcal{N}}\int d\mathbf{x}_1\cdots d\mathbf{x}_N\Psi(\mathbf{x}_1,\cdots ,\mathbf{x}_N)|\mathbf{x}_1, \cdots ,\mathbf{x}_N\rangle$ where the first $N/2$ coordinates refer to spin up atoms and the second to spin down, we have
\begin{multline}
I(\boldsymbol\kappa)=\frac{1}{\mathcal{N}}\int d\mathbf{x}_1\cdots d\mathbf{x}_N|\Psi(\mathbf{x}_1,\cdots ,\mathbf{x}_N)|^2\\
\times\left|\sum_{j=1}^{N/2}\left(e^{-\boldsymbol\kappa\cdot\mathbf{x}_j}-
e^{-\boldsymbol\kappa\cdot\mathbf{x}_{j+N/2}}\right)\right|^2.
\label{eq:Far_field_integral}
\end{multline}
The far-field images for the states from Eqs.~(\ref{sqhs_ground_state})-(\ref{eq:pfaffian_state}) can then be calculated by using the Metropolis Monte Carlo method to evaluate the integral over $N$ coordinates (see, e.g., Ref \cite{Newman1999a}).

In Figs.~\ref{sqs_far_field}(a) and \ref{sqs_far_field}(b) we show the far field image of the bosonic Halperin state from Eq.~(\ref{sqhs_ground_state}) and the Pfaffian state from Eq.~(\ref{eq:pfaffian_state}). For comparison in Fig.~\ref{sqs_far_field}(c) we show the image of a state with $N/2$ atoms of each spin but with the spatial wave function given by the Laughlin wave function in Eq.~(\ref{eq:Laughlin_state}) with $n=2$. The images are generated by averaging over 100000 Monte Carlo configurations of 100 particles. 
The far-field images allow us to distinguish the three different states, where all the images go to zero at $\boldsymbol\kappa = \mathbf{0}$ but the nature of this approach varies for each state as shown in Fig.~\ref{sqs_far_field}(d). The difference is due to the different spatial correlations between opposite spins in each wave function. In the Halperin state the atoms of opposite spin are allowed to exist closer to each other than particles of the same spin, which results from the lower power of the $(z_j-z_l)$ terms for atoms of opposite spin in the wave function. This leads to long wavelength oscillations in the far field that create the large dip in the image of this state. For the Laughlin state the particles of opposite spin interact in the same way as particles of the same spin and these long-wavelength oscillations do not occur. In the Pfaffian state the $(z_j-z_l)$ term does not enter the wave function for atoms in a pair, and paired atoms come much closer to one another than other atoms leading to an intermediate dip at $\boldsymbol\kappa = \mathbf{0}$.

\begin{figure}
\centering
\includegraphics{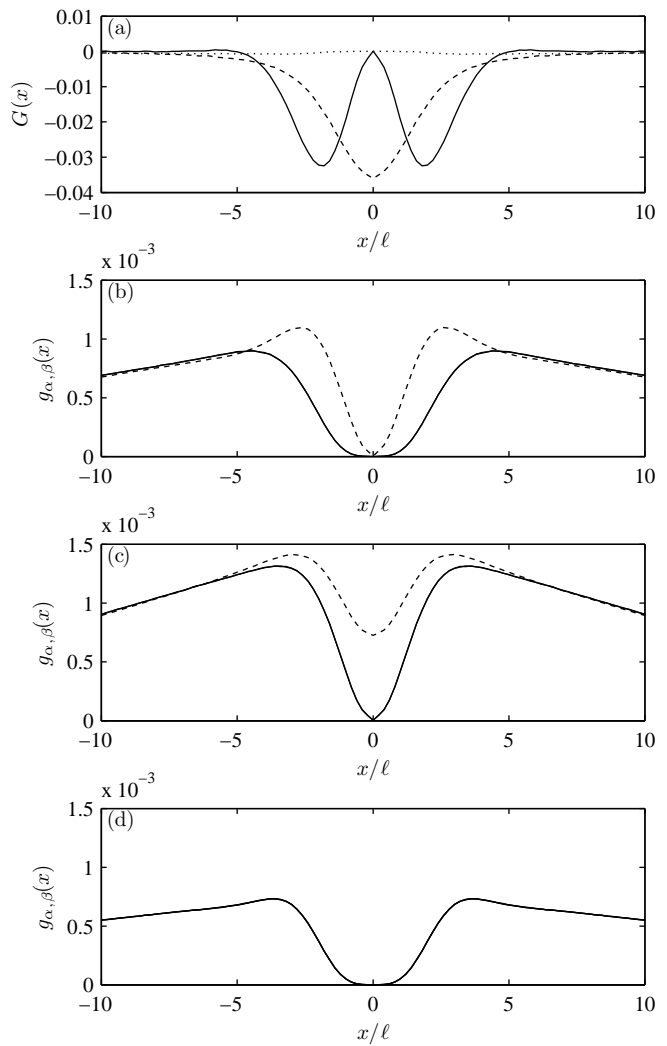}
\caption{(a) Slice along $y=0$ of the transform given in Eq.~(\ref{eq:Image_transform}) of the far-field images  for the Halperin state (solid), the Pfaffian state (dashed) and the comparison Laughlin state (dotted). [(b)-(d)] Two particle correlation functions for the (b) Halperin state, (c) the Pfaffian state, and (d) the Laughlin state, with  $g_{++}(x,y=0)=g_{--}(x,y=0)$ (solid lines) and $g_{+-}(x,y=0)=g_{-+}(x,y=0)$ (dashed lines). Functions were calculated using 100000 Monte Carlo configurations of 100 particles.}
\label{fig:image_transform}
\end{figure}

In fact, the far-field image of the gas is related to its two-particle spin-correlation function by Fourier transform. We have
\begin{align}
G(\mathbf{x}) =& \int \frac{d\boldsymbol\kappa}{(2\pi)^2}\left(\frac{I(\boldsymbol\kappa)}{N}-1\right)e^{i \boldsymbol\kappa\cdot\mathbf{x}} \notag\\
=&-\frac{N}{4}(g_{\uparrow\downarrow}(\mathbf{x})+g_{\downarrow\uparrow}(\mathbf{x}))\label{eq:Image_transform}\\&+\frac{\left(N/2-1\right)}{2}\left(g_{\uparrow\uparrow}(\mathbf{x})+g_{\downarrow\downarrow}(\mathbf{x})\right)\notag
\end{align}
where
\begin{equation}
g_{\alpha,\beta}(\mathbf{x}) = \int d\mathbf{x}'\langle \hat{\Psi}^\dagger_{\beta}(\mathbf{x}')\hat{\Psi}^\dagger_{\alpha}(\mathbf{x}'+\mathbf{x})\hat{\Psi}_{\alpha}(\mathbf{x}'+\mathbf{x})\hat{\Psi}_{\beta}(\mathbf{x}')\rangle
\end{equation}
are the two-particle correlation functions, which give the probability of finding a particle in state $\alpha$ separated by $\mathbf{x}$ from a particle in state $\beta$. In Fig.~\ref{fig:image_transform}(a) we plot a slice through the transform of the far-field images given by Eq.~(\ref{eq:Image_transform}). In Figs.~\ref{fig:image_transform}(b)-\ref{fig:image_transform}(d) we show the corresponding correlation functions for the Halperin state, the Pfaffian state, and our comparison Laughlin state. The combination of the correlation functions given in Eq.~(\ref{eq:Image_transform}) leads largely to cancellation in the Laughlin case as all the correlation functions are equal. In the Halperin case $g_{\downarrow\uparrow}(\mathbf{x})$ exhibits extra lobes at $x\sim 2.5\ell$ that do not cancel in the sum and show up in the transformed image, similarly in the Pfaffian case $g_{\downarrow\uparrow}(\mathbf{x})$ is non-zero close to $\mathbf{x}=0$ which does not cancel with the other terms.


In Figs.~\ref{sqs_far_field} and \ref{fig:image_transform} we have averaged over 100000 Monte Carlo configurations to estimate the expectation value in Eq.~(\ref{eq:Far_field_integral}). In a real experiment an image from a single experimental run will not give the expectation value, the combination of scattering and detection of photons will disturb the state leading to relative localization of the atoms \cite{Douglas2011b}, and the image will be determined by the equivalent of a single Monte Carlo configuration. This does not mean 100000 experimental runs are required to distinguish the states; in fact, images formed by averaging over 10 configurations, although noisy, show the characteristic differences between the states allowing the states to be distinguished in a realistic number of experimental runs. Furthermore, because of the radial symmetry of the many-body state finding the radial intensity of the image by averaging over all angles can improve the ability to distinguish the states for a small number of experimental runs.

Another experimental issue is the effect of the trapping potential used to trap the ultracold gas on the many-body state. For optically induced gauge potentials the trap potential and optically induced scalar potential can be arranged to cancel \cite{Klein2005a}. For rotating gases in the high rotation limit the trapping potential is a crucial factor in realizing the effective Hamiltonian of a magnetic field, and at the critical rotation frequency the trapping potential plays no additional role. In both these experiment arrangements the many-body states of the system are expected to be the ones considered above \cite{Paredes2002a}; however, if an excess trapping potential remains in addition to the magnetic-field Hamiltonian, the state can form domains of quantum Hall states with different filling factors that would lead to an altered far-field image \cite{Cooper2005a}. We do not consider this case here.


\section{Images of quasiholes}

It is also interesting to consider what imaging can reveal about quasiparticles in these quantum Hall systems. For multicomponent systems, quasiholes can be created by a focused laser which repels atoms at a particular position, say, $\eta$. The laser can be arranged to primarily affect only one of the spin components or, alternatively, multiple components \cite{Paredes2002a}. For the two-component bosonic Halperin state where the the laser affects only the spin-up atoms the state is then given by
\begin{equation}
\Psi_\eta  = S\left[\prod_{i=1}^{N/2} (z_i-\eta) \psi\right].
\label{eq:two_comp_quasihole}
\end{equation}
In effect, the spin-up particles see an extra spin-down particle at $\eta$, leading to a repulsion from $\eta$ and a lower number of spin-up particles in the region close to $\eta$. The spin-down particles, on the other hand, are not affected by the laser and a higher than normal density of spin-down particles will exist at $\eta$. This type of quasihole is an anyon with $1/3$ statistics.


\begin{figure}
\centering
\includegraphics{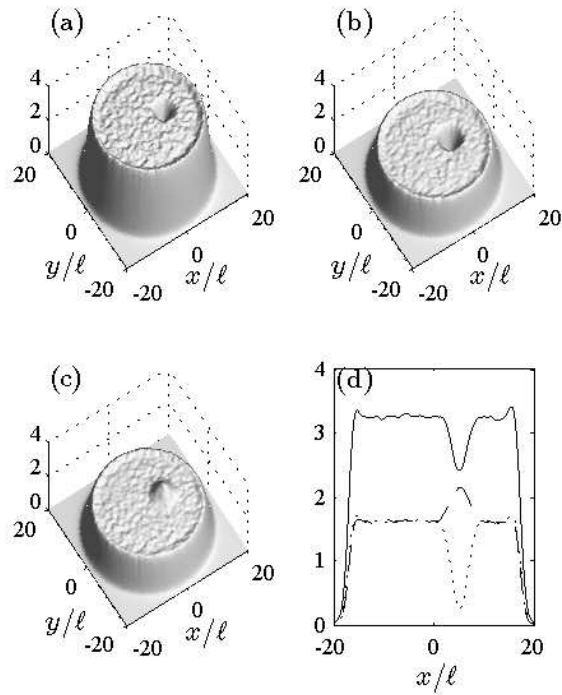}
\caption{Images of the two-component Halperin state with a spin-up quasihole at $(x/\ell,y/\ell) =(5,0)$. (a) Total density, (b) spin-up density, and (c) spin-down density. (d) Slice with $y/\ell = 0$ through images of the (solid) total density, spin-up density, (dotted) and spin-down density (dashed). Images were calculated using 100000 Monte Carlo configurations of 100 particles for an imaging system with numerical aperture $\sin \theta = 0.5\lambda/\ell$. Images are normalized to the total number of particles.}
\label{fig:two_comp_quasihole}
\end{figure}


Far-field images of states with a small number of quasiholes do not differ significantly from the corresponding image of the state without quasiholes. This is because adding a quasihole causes only a local defect to the state and hence does not change the particle-particle correlation function significantly.  Instead, microscopic imaging can be used to reveal the density of the state \cite{Douglas2010a}. In this case, light scattered by the sample is collected by a microscope with numerical aperture $\sin \theta$ and forms an image of the density or spin density. We can choose the probe light so that it scatters identically from either spin state or so that it scatters from only one state, giving a total density image or an image of the density of either spin state.

Images of a bosonic Halperin state with a spin-up quasihole at $(x/\ell,y/\ell) =(5,0)$ are shown in Fig.~\ref{fig:two_comp_quasihole}. We can see a dip in the total density and a larger than normal density of spin-down atoms at the quasihole position. In fact, the quasihole statistics are revealed from these images by integrating the image intensity over the region of excess or reduction to find the volume added to or removed from the normal density. We find that the dip in the total density has a volume of 1/3 of an atom, and the peak in the spin-down atom density also has a volume of 1/3 of an atom. The dip in the spin-up density has a volume of 2/3 of an atom. This property is also seen in the images of similar states with more than two spin components. For the $n$-component generalization of the Halperin state, the quasiparticles have $p/(1+n)$ statistics, where $p$ is the number of components the quasihole laser affects \cite{Paredes2002a}. Then, for example, in images of a three-component gas where two components are affected by the laser, the total density image has a dip of volume 1/2 of an atom at $\eta$, while the unaffected spin component has a bump of volume 1/2 of an atom at $\eta$. In general, the total density and the unaffected components will have less and excess, respectively, of $p/(n+1)$ of an atom at $\eta$ and the affected components will have $1-p/(n+1)$ less of an atom at $\eta$.


\section{Summary}

The above examples demonstrate the potential that imaging of scattered light has as a tool for investigating the properties of ultracold atomic gases in the quantum Hall regime. Imaging in the far field provides a probe of the density-density correlation of the atomic matter and when the light scattering is spin dependent it provides a probe of the spin-density correlations. Different states then have different optical signatures that relate directly to the spin-correlation functions of the state, enabling the identification of particular states in experiment. Furthermore, imaging the density using a microscope can reveal quasiholes, where the reduction in the density at the quasihole position gives the quasihole statistics.


%


\begin{thebibliography}{44}%
\makeatletter
\providecommand \@ifxundefined [1]{%
 \@ifx{#1\undefined}
}%
\providecommand \@ifnum [1]{%
 \ifnum #1\expandafter \@firstoftwo
 \else \expandafter \@secondoftwo
 \fi
}%
\providecommand \@ifx [1]{%
 \ifx #1\expandafter \@firstoftwo
 \else \expandafter \@secondoftwo
 \fi
}%
\providecommand \natexlab [1]{#1}%
\providecommand \enquote  [1]{``#1''}%
\providecommand \bibnamefont  [1]{#1}%
\providecommand \bibfnamefont [1]{#1}%
\providecommand \citenamefont [1]{#1}%
\providecommand \href@noop [0]{\@secondoftwo}%
\providecommand \href [0]{\begingroup \@sanitize@url \@href}%
\providecommand \@href[1]{\@@startlink{#1}\@@href}%
\providecommand \@@href[1]{\endgroup#1\@@endlink}%
\providecommand \@sanitize@url [0]{\catcode `\\12\catcode `\$12\catcode
  `\&12\catcode `\#12\catcode `\^12\catcode `\_12\catcode `\%12\relax}%
\providecommand \@@startlink[1]{}%
\providecommand \@@endlink[0]{}%
\providecommand \url  [0]{\begingroup\@sanitize@url \@url }%
\providecommand \@url [1]{\endgroup\@href {#1}{\urlprefix }}%
\providecommand \urlprefix  [0]{URL }%
\providecommand \Eprint [0]{\href }%
\providecommand \doibase [0]{http://dx.doi.org/}%
\providecommand \selectlanguage [0]{\@gobble}%
\providecommand \bibinfo  [0]{\@secondoftwo}%
\providecommand \bibfield  [0]{\@secondoftwo}%
\providecommand \translation [1]{[#1]}%
\providecommand \BibitemOpen [0]{}%
\providecommand \bibitemStop [0]{}%
\providecommand \bibitemNoStop [0]{.\EOS\space}%
\providecommand \EOS [0]{\spacefactor3000\relax}%
\providecommand \BibitemShut  [1]{\csname bibitem#1\endcsname}%
\let\auto@bib@innerbib\@empty
\bibitem [{\citenamefont {Klitzing}\ \emph {et~al.}(1980)\citenamefont
  {Klitzing}, \citenamefont {Dorda},\ and\ \citenamefont
  {Pepper}}]{Klitzing1980a}%
  \BibitemOpen
  \bibfield  {author} {\bibinfo {author} {\bibfnamefont {K.~v.}\ \bibnamefont
  {Klitzing}}, \bibinfo {author} {\bibfnamefont {G.}~\bibnamefont {Dorda}}, \
  and\ \bibinfo {author} {\bibfnamefont {M.}~\bibnamefont {Pepper}},\ }\href
  {\doibase 10.1103/PhysRevLett.45.494} {\bibfield  {journal} {\bibinfo
  {journal} {Phys. Rev. Lett.}\ }\textbf {\bibinfo {volume} {45}},\ \bibinfo
  {pages} {494} (\bibinfo {year} {1980})}\BibitemShut {NoStop}%
\bibitem [{\citenamefont {Tsui}\ \emph {et~al.}(1982)\citenamefont {Tsui},
  \citenamefont {Stormer},\ and\ \citenamefont {Gossard}}]{Tsui1982a}%
  \BibitemOpen
  \bibfield  {author} {\bibinfo {author} {\bibfnamefont {D.~C.}\ \bibnamefont
  {Tsui}}, \bibinfo {author} {\bibfnamefont {H.~L.}\ \bibnamefont {Stormer}}, \
  and\ \bibinfo {author} {\bibfnamefont {A.~C.}\ \bibnamefont {Gossard}},\
  }\href {\doibase 10.1103/PhysRevLett.48.1559} {\bibfield  {journal} {\bibinfo
   {journal} {Phys. Rev. Lett.}\ }\textbf {\bibinfo {volume} {48}},\ \bibinfo
  {pages} {1559} (\bibinfo {year} {1982})}\BibitemShut {NoStop}%
\bibitem [{\citenamefont {Laughlin}(1983)}]{Laughlin1983a}%
  \BibitemOpen
  \bibfield  {author} {\bibinfo {author} {\bibfnamefont {R.~B.}\ \bibnamefont
  {Laughlin}},\ }\href {\doibase 10.1103/PhysRevLett.50.1395} {\bibfield
  {journal} {\bibinfo  {journal} {Phys. Rev. Lett.}\ }\textbf {\bibinfo
  {volume} {50}},\ \bibinfo {pages} {1395} (\bibinfo {year}
  {1983})}\BibitemShut {NoStop}%
\bibitem [{\citenamefont {Halperin}(1983)}]{Halperin1983a}%
  \BibitemOpen
  \bibfield  {author} {\bibinfo {author} {\bibfnamefont {B.~I.}\ \bibnamefont
  {Halperin}},\ }\href@noop {} {\bibfield  {journal} {\bibinfo  {journal}
  {Helv. Phys. Acta}\ }\textbf {\bibinfo {volume} {56}},\ \bibinfo {pages} {75}
  (\bibinfo {year} {1983})}\BibitemShut {NoStop}%
\bibitem [{\citenamefont {Chakraborty}\ and\ \citenamefont
  {Zhang}(1984)}]{Chakraborty1984a}%
  \BibitemOpen
  \bibfield  {author} {\bibinfo {author} {\bibfnamefont {T.}~\bibnamefont
  {Chakraborty}}\ and\ \bibinfo {author} {\bibfnamefont {F.~C.}\ \bibnamefont
  {Zhang}},\ }\href {\doibase 10.1103/PhysRevB.29.7032} {\bibfield  {journal}
  {\bibinfo  {journal} {Phys. Rev. B}\ }\textbf {\bibinfo {volume} {29}},\
  \bibinfo {pages} {7032} (\bibinfo {year} {1984})}\BibitemShut {NoStop}%
\bibitem [{\citenamefont {Rezayi}(1987)}]{Rezayi1987a}%
  \BibitemOpen
  \bibfield  {author} {\bibinfo {author} {\bibfnamefont {E.~H.}\ \bibnamefont
  {Rezayi}},\ }\href {\doibase 10.1103/PhysRevB.36.5454} {\bibfield  {journal}
  {\bibinfo  {journal} {Phys. Rev. B}\ }\textbf {\bibinfo {volume} {36}},\
  \bibinfo {pages} {5454} (\bibinfo {year} {1987})}\BibitemShut {NoStop}%
\bibitem [{\citenamefont {Bloch}\ \emph {et~al.}(2008)\citenamefont {Bloch},
  \citenamefont {Dalibard},\ and\ \citenamefont {Zwerger}}]{Bloch2008a}%
  \BibitemOpen
  \bibfield  {author} {\bibinfo {author} {\bibfnamefont {I.}~\bibnamefont
  {Bloch}}, \bibinfo {author} {\bibfnamefont {J.}~\bibnamefont {Dalibard}}, \
  and\ \bibinfo {author} {\bibfnamefont {W.}~\bibnamefont {Zwerger}},\ }\href
  {\doibase 10.1103/RevModPhys.80.885} {\bibfield  {journal} {\bibinfo
  {journal} {Rev. Mod. Phys.}\ }\textbf {\bibinfo {volume} {80}},\ \bibinfo
  {pages} {885} (\bibinfo {year} {2008})}\BibitemShut {NoStop}%
\bibitem [{\citenamefont {Lewenstein}\ \emph {et~al.}(2007)\citenamefont
  {Lewenstein}, \citenamefont {Sanpera}, \citenamefont {Ahufinger},
  \citenamefont {Damski}, \citenamefont {De},\ and\ \citenamefont
  {Sen}}]{Lewenstein2007a}%
  \BibitemOpen
  \bibfield  {author} {\bibinfo {author} {\bibfnamefont {M.}~\bibnamefont
  {Lewenstein}}, \bibinfo {author} {\bibfnamefont {A.}~\bibnamefont {Sanpera}},
  \bibinfo {author} {\bibfnamefont {V.}~\bibnamefont {Ahufinger}}, \bibinfo
  {author} {\bibfnamefont {B.}~\bibnamefont {Damski}}, \bibinfo {author}
  {\bibfnamefont {A.~S.}\ \bibnamefont {De}}, \ and\ \bibinfo {author}
  {\bibfnamefont {U.}~\bibnamefont {Sen}},\ }\href {\doibase
  doi:10.1080/00018730701223200} {\bibfield  {journal} {\bibinfo  {journal}
  {Adv. Phys.}\ }\textbf {\bibinfo {volume} {56}},\ \bibinfo {pages} {243}
  (\bibinfo {year} {March 2007})}\BibitemShut {NoStop}%
\bibitem [{\citenamefont {Cooper}(2008)}]{Cooper2008a}%
  \BibitemOpen
  \bibfield  {author} {\bibinfo {author} {\bibfnamefont {N.~R.}\ \bibnamefont
  {Cooper}},\ }\href {http://www.informaworld.com/10.1080/00018730802564122}
  {\bibfield  {journal} {\bibinfo  {journal} {Advances in Physics}\ }\textbf
  {\bibinfo {volume} {57}},\ \bibinfo {pages} {539} (\bibinfo {year}
  {2008})}\BibitemShut {NoStop}%
\bibitem [{\citenamefont {Lin}\ \emph {et~al.}(2009)\citenamefont {Lin},
  \citenamefont {Compton}, \citenamefont {Jimenez-Garcia}, \citenamefont
  {Porto},\ and\ \citenamefont {Spielman}}]{Lin2009a}%
  \BibitemOpen
  \bibfield  {author} {\bibinfo {author} {\bibfnamefont {Y.-J.}\ \bibnamefont
  {Lin}}, \bibinfo {author} {\bibfnamefont {R.~L.}\ \bibnamefont {Compton}},
  \bibinfo {author} {\bibfnamefont {K.}~\bibnamefont {Jimenez-Garcia}},
  \bibinfo {author} {\bibfnamefont {J.~V.}\ \bibnamefont {Porto}}, \ and\
  \bibinfo {author} {\bibfnamefont {I.~B.}\ \bibnamefont {Spielman}},\ }\href
  {http://dx.doi.org/10.1038/nature08609} {\bibfield  {journal} {\bibinfo
  {journal} {Nature}\ }\textbf {\bibinfo {volume} {462}},\ \bibinfo {pages}
  {628} (\bibinfo {year} {2009})}\BibitemShut {NoStop}%
\bibitem [{\citenamefont {{Dalibard}}\ \emph {et~al.}(2010)\citenamefont
  {{Dalibard}}, \citenamefont {{Gerbier}}, \citenamefont {{Juzeli{\= u}nas}},\
  and\ \citenamefont {{{\"O}hberg}}}]{Dalibard2010a}%
  \BibitemOpen
  \bibfield  {author} {\bibinfo {author} {\bibfnamefont {J.}~\bibnamefont
  {{Dalibard}}}, \bibinfo {author} {\bibfnamefont {F.}~\bibnamefont
  {{Gerbier}}}, \bibinfo {author} {\bibfnamefont {G.}~\bibnamefont {{Juzeli{\=
  u}nas}}}, \ and\ \bibinfo {author} {\bibfnamefont {P.}~\bibnamefont
  {{{\"O}hberg}}},\ }\href@noop {} {\bibfield  {journal} {\bibinfo  {journal}
  {ArXiv e-prints}\ } (\bibinfo {year} {2010})},\ \Eprint
  {http://arxiv.org/abs/1008.5378} {arXiv:1008.5378 [cond-mat.quant-gas]}
  \BibitemShut {NoStop}%
\bibitem [{\citenamefont {Nayak}\ \emph {et~al.}(2008)\citenamefont {Nayak},
  \citenamefont {Simon}, \citenamefont {Stern}, \citenamefont {Freedman},\ and\
  \citenamefont {Das~Sarma}}]{Nayak2008a}%
  \BibitemOpen
  \bibfield  {author} {\bibinfo {author} {\bibfnamefont {C.}~\bibnamefont
  {Nayak}}, \bibinfo {author} {\bibfnamefont {S.~H.}\ \bibnamefont {Simon}},
  \bibinfo {author} {\bibfnamefont {A.}~\bibnamefont {Stern}}, \bibinfo
  {author} {\bibfnamefont {M.}~\bibnamefont {Freedman}}, \ and\ \bibinfo
  {author} {\bibfnamefont {S.}~\bibnamefont {Das~Sarma}},\ }\href {\doibase
  10.1103/RevModPhys.80.1083} {\bibfield  {journal} {\bibinfo  {journal} {Rev.
  Mod. Phys.}\ }\textbf {\bibinfo {volume} {80}},\ \bibinfo {pages} {1083}
  (\bibinfo {year} {2008})}\BibitemShut {NoStop}%
\bibitem [{\citenamefont {Read}\ and\ \citenamefont
  {Cooper}(2003)}]{Read2003a}%
  \BibitemOpen
  \bibfield  {author} {\bibinfo {author} {\bibfnamefont {N.}~\bibnamefont
  {Read}}\ and\ \bibinfo {author} {\bibfnamefont {N.~R.}\ \bibnamefont
  {Cooper}},\ }\href {\doibase 10.1103/PhysRevA.68.035601} {\bibfield
  {journal} {\bibinfo  {journal} {Phys. Rev. A}\ }\textbf {\bibinfo {volume}
  {68}},\ \bibinfo {pages} {035601} (\bibinfo {year} {2003})}\BibitemShut
  {NoStop}%
\bibitem [{\citenamefont {S\o{}rensen}\ \emph {et~al.}(2005)\citenamefont
  {S\o{}rensen}, \citenamefont {Demler},\ and\ \citenamefont
  {Lukin}}]{Lukin2005a}%
  \BibitemOpen
  \bibfield  {author} {\bibinfo {author} {\bibfnamefont {A.~S.}\ \bibnamefont
  {S\o{}rensen}}, \bibinfo {author} {\bibfnamefont {E.}~\bibnamefont {Demler}},
  \ and\ \bibinfo {author} {\bibfnamefont {M.~D.}\ \bibnamefont {Lukin}},\
  }\href {\doibase 10.1103/PhysRevLett.94.086803} {\bibfield  {journal}
  {\bibinfo  {journal} {Phys. Rev. Lett.}\ }\textbf {\bibinfo {volume} {94}},\
  \bibinfo {pages} {086803} (\bibinfo {year} {2005})}\BibitemShut {NoStop}%
\bibitem [{\citenamefont {Palmer}\ \emph {et~al.}(2008)\citenamefont {Palmer},
  \citenamefont {Klein},\ and\ \citenamefont {Jaksch}}]{Palmer2008a}%
  \BibitemOpen
  \bibfield  {author} {\bibinfo {author} {\bibfnamefont {R.~N.}\ \bibnamefont
  {Palmer}}, \bibinfo {author} {\bibfnamefont {A.}~\bibnamefont {Klein}}, \
  and\ \bibinfo {author} {\bibfnamefont {D.}~\bibnamefont {Jaksch}},\ }\href
  {\doibase 10.1103/PhysRevA.78.013609} {\bibfield  {journal} {\bibinfo
  {journal} {Phys. Rev. A}\ }\textbf {\bibinfo {volume} {78}},\ \bibinfo
  {pages} {013609} (\bibinfo {year} {2008})}\BibitemShut {NoStop}%
\bibitem [{\citenamefont {Cooper}\ \emph {et~al.}(2005)\citenamefont {Cooper},
  \citenamefont {van Lankvelt}, \citenamefont {Reijnders},\ and\ \citenamefont
  {Schoutens}}]{Cooper2005a}%
  \BibitemOpen
  \bibfield  {author} {\bibinfo {author} {\bibfnamefont {N.~R.}\ \bibnamefont
  {Cooper}}, \bibinfo {author} {\bibfnamefont {F.~J.~M.}\ \bibnamefont {van
  Lankvelt}}, \bibinfo {author} {\bibfnamefont {J.~W.}\ \bibnamefont
  {Reijnders}}, \ and\ \bibinfo {author} {\bibfnamefont {K.}~\bibnamefont
  {Schoutens}},\ }\href {\doibase 10.1103/PhysRevA.72.063622} {\bibfield
  {journal} {\bibinfo  {journal} {Phys. Rev. A}\ }\textbf {\bibinfo {volume}
  {72}},\ \bibinfo {pages} {063622} (\bibinfo {year} {2005})}\BibitemShut
  {NoStop}%
\bibitem [{\citenamefont {Cazalilla}\ \emph {et~al.}(2005)\citenamefont
  {Cazalilla}, \citenamefont {Barber\'an},\ and\ \citenamefont
  {Cooper}}]{Cazalilla2005a}%
  \BibitemOpen
  \bibfield  {author} {\bibinfo {author} {\bibfnamefont {M.~A.}\ \bibnamefont
  {Cazalilla}}, \bibinfo {author} {\bibfnamefont {N.}~\bibnamefont
  {Barber\'an}}, \ and\ \bibinfo {author} {\bibfnamefont {N.~R.}\ \bibnamefont
  {Cooper}},\ }\href {\doibase 10.1103/PhysRevB.71.121303} {\bibfield
  {journal} {\bibinfo  {journal} {Phys. Rev. B}\ }\textbf {\bibinfo {volume}
  {71}},\ \bibinfo {pages} {121303} (\bibinfo {year} {2005})}\BibitemShut
  {NoStop}%
\bibitem [{\citenamefont {Stanescu}\ \emph {et~al.}(2010)\citenamefont
  {Stanescu}, \citenamefont {Galitski},\ and\ \citenamefont
  {Das~Sarma}}]{Stanescu2010a}%
  \BibitemOpen
  \bibfield  {author} {\bibinfo {author} {\bibfnamefont {T.~D.}\ \bibnamefont
  {Stanescu}}, \bibinfo {author} {\bibfnamefont {V.}~\bibnamefont {Galitski}},
  \ and\ \bibinfo {author} {\bibfnamefont {S.}~\bibnamefont {Das~Sarma}},\
  }\href {\doibase 10.1103/PhysRevA.82.013608} {\bibfield  {journal} {\bibinfo
  {journal} {Phys. Rev. A}\ }\textbf {\bibinfo {volume} {82}},\ \bibinfo
  {pages} {013608} (\bibinfo {year} {2010})}\BibitemShut {NoStop}%
\bibitem [{\citenamefont {Hafezi}\ \emph {et~al.}(2007)\citenamefont {Hafezi},
  \citenamefont {S{\o}rensen}, \citenamefont {Demler},\ and\ \citenamefont
  {Lukin}}]{Hafezi2007a}%
  \BibitemOpen
  \bibfield  {author} {\bibinfo {author} {\bibfnamefont {M.}~\bibnamefont
  {Hafezi}}, \bibinfo {author} {\bibfnamefont {A.~S.}\ \bibnamefont
  {S{\o}rensen}}, \bibinfo {author} {\bibfnamefont {E.}~\bibnamefont {Demler}},
  \ and\ \bibinfo {author} {\bibfnamefont {M.~D.}\ \bibnamefont {Lukin}},\
  }\href {\doibase 10.1103/PhysRevA.76.023613} {\bibfield  {journal} {\bibinfo
  {journal} {Phys. Rev. A}\ }\textbf {\bibinfo {volume} {76}},\ \bibinfo {eid}
  {023613} (\bibinfo {year} {2007})}\BibitemShut {NoStop}%
\bibitem [{\citenamefont {Liu}\ \emph {et~al.}(2010{\natexlab{a}})\citenamefont
  {Liu}, \citenamefont {Liu}, \citenamefont {Wu},\ and\ \citenamefont
  {Sinova}}]{Liu2010b}%
  \BibitemOpen
  \bibfield  {author} {\bibinfo {author} {\bibfnamefont {X.-J.}\ \bibnamefont
  {Liu}}, \bibinfo {author} {\bibfnamefont {X.}~\bibnamefont {Liu}}, \bibinfo
  {author} {\bibfnamefont {C.}~\bibnamefont {Wu}}, \ and\ \bibinfo {author}
  {\bibfnamefont {J.}~\bibnamefont {Sinova}},\ }\href {\doibase
  10.1103/PhysRevA.81.033622} {\bibfield  {journal} {\bibinfo  {journal} {Phys.
  Rev. A}\ }\textbf {\bibinfo {volume} {81}},\ \bibinfo {pages} {033622}
  (\bibinfo {year} {2010}{\natexlab{a}})}\BibitemShut {NoStop}%
\bibitem [{\citenamefont {Shao}\ \emph {et~al.}(2008)\citenamefont {Shao},
  \citenamefont {Zhu}, \citenamefont {Sheng}, \citenamefont {Xing},\ and\
  \citenamefont {Wang}}]{Shao2008a}%
  \BibitemOpen
  \bibfield  {author} {\bibinfo {author} {\bibfnamefont {L.~B.}\ \bibnamefont
  {Shao}}, \bibinfo {author} {\bibfnamefont {S.-L.}\ \bibnamefont {Zhu}},
  \bibinfo {author} {\bibfnamefont {L.}~\bibnamefont {Sheng}}, \bibinfo
  {author} {\bibfnamefont {D.~Y.}\ \bibnamefont {Xing}}, \ and\ \bibinfo
  {author} {\bibfnamefont {Z.~D.}\ \bibnamefont {Wang}},\ }\href {\doibase
  10.1103/PhysRevLett.101.246810} {\bibfield  {journal} {\bibinfo  {journal}
  {Phys. Rev. Lett.}\ }\textbf {\bibinfo {volume} {101}},\ \bibinfo {pages}
  {246810} (\bibinfo {year} {2008})}\BibitemShut {NoStop}%
\bibitem [{\citenamefont {Liu}\ \emph {et~al.}(2010{\natexlab{b}})\citenamefont
  {Liu}, \citenamefont {Zhu}, \citenamefont {Jiang}, \citenamefont {Sun},\ and\
  \citenamefont {Liu}}]{Liu2010a}%
  \BibitemOpen
  \bibfield  {author} {\bibinfo {author} {\bibfnamefont {G.}~\bibnamefont
  {Liu}}, \bibinfo {author} {\bibfnamefont {S.-L.}\ \bibnamefont {Zhu}},
  \bibinfo {author} {\bibfnamefont {S.}~\bibnamefont {Jiang}}, \bibinfo
  {author} {\bibfnamefont {F.}~\bibnamefont {Sun}}, \ and\ \bibinfo {author}
  {\bibfnamefont {W.~M.}\ \bibnamefont {Liu}},\ }\href {\doibase
  10.1103/PhysRevA.82.053605} {\bibfield  {journal} {\bibinfo  {journal} {Phys.
  Rev. A}\ }\textbf {\bibinfo {volume} {82}},\ \bibinfo {pages} {053605}
  (\bibinfo {year} {2010}{\natexlab{b}})}\BibitemShut {NoStop}%
\bibitem [{\citenamefont {Umucal{\i}lar}\ \emph {et~al.}(2008)\citenamefont
  {Umucal{\i}lar}, \citenamefont {Zhai},\ and\ \citenamefont
  {Oktel}}]{Oktel2008a}%
  \BibitemOpen
  \bibfield  {author} {\bibinfo {author} {\bibfnamefont {R.~O.}\ \bibnamefont
  {Umucal{\i}lar}}, \bibinfo {author} {\bibfnamefont {H.}~\bibnamefont {Zhai}},
  \ and\ \bibinfo {author} {\bibfnamefont {M.~O.}\ \bibnamefont {Oktel}},\
  }\href {\doibase 10.1103/PhysRevLett.100.070402} {\bibfield  {journal}
  {\bibinfo  {journal} {Phys. Rev. Lett.}\ }\textbf {\bibinfo {volume} {100}},\
  \bibinfo {pages} {070402} (\bibinfo {year} {2008})}\BibitemShut {NoStop}%
\bibitem [{\citenamefont {Paredes}\ \emph {et~al.}(2001)\citenamefont
  {Paredes}, \citenamefont {Fedichev}, \citenamefont {Cirac},\ and\
  \citenamefont {Zoller}}]{Paredes2001a}%
  \BibitemOpen
  \bibfield  {author} {\bibinfo {author} {\bibfnamefont {B.}~\bibnamefont
  {Paredes}}, \bibinfo {author} {\bibfnamefont {P.~O.}\ \bibnamefont
  {Fedichev}}, \bibinfo {author} {\bibfnamefont {J.~I.}\ \bibnamefont {Cirac}},
  \ and\ \bibinfo {author} {\bibfnamefont {P.}~\bibnamefont {Zoller}},\
  }\href@noop {} {\bibfield  {journal} {\bibinfo  {journal} {Phys. Rev. Lett.}\
  }\textbf {\bibinfo {volume} {87}},\ \bibinfo {pages} {010402} (\bibinfo
  {year} {2001})}\BibitemShut {NoStop}%
\bibitem [{\citenamefont {Paredes}\ \emph {et~al.}(2002)\citenamefont
  {Paredes}, \citenamefont {Zoller},\ and\ \citenamefont
  {Cirac}}]{Paredes2002a}%
  \BibitemOpen
  \bibfield  {author} {\bibinfo {author} {\bibfnamefont {B.}~\bibnamefont
  {Paredes}}, \bibinfo {author} {\bibfnamefont {P.}~\bibnamefont {Zoller}}, \
  and\ \bibinfo {author} {\bibfnamefont {J.~I.}\ \bibnamefont {Cirac}},\ }\href
  {\doibase 10.1103/PhysRevA.66.033609} {\bibfield  {journal} {\bibinfo
  {journal} {Phys. Rev. A}\ }\textbf {\bibinfo {volume} {66}},\ \bibinfo
  {pages} {033609} (\bibinfo {year} {2002})}\BibitemShut {NoStop}%
\bibitem [{\citenamefont {Juzeli\ifmmode~\bar{u}\else \={u}\fi{}nas}\ \emph
  {et~al.}(2005)\citenamefont {Juzeli\ifmmode~\bar{u}\else \={u}\fi{}nas},
  \citenamefont {\"Ohberg}, \citenamefont {Ruseckas},\ and\ \citenamefont
  {Klein}}]{Klein2005a}%
  \BibitemOpen
  \bibfield  {author} {\bibinfo {author} {\bibfnamefont {G.}~\bibnamefont
  {Juzeli\ifmmode~\bar{u}\else \={u}\fi{}nas}}, \bibinfo {author}
  {\bibfnamefont {P.}~\bibnamefont {\"Ohberg}}, \bibinfo {author}
  {\bibfnamefont {J.}~\bibnamefont {Ruseckas}}, \ and\ \bibinfo {author}
  {\bibfnamefont {A.}~\bibnamefont {Klein}},\ }\href {\doibase
  10.1103/PhysRevA.71.053614} {\bibfield  {journal} {\bibinfo  {journal} {Phys.
  Rev. A}\ }\textbf {\bibinfo {volume} {71}},\ \bibinfo {pages} {053614}
  (\bibinfo {year} {2005})}\BibitemShut {NoStop}%
\bibitem [{\citenamefont {Moore}\ and\ \citenamefont
  {Read}(1991)}]{Moore1991a}%
  \BibitemOpen
  \bibfield  {author} {\bibinfo {author} {\bibfnamefont {G.}~\bibnamefont
  {Moore}}\ and\ \bibinfo {author} {\bibfnamefont {N.}~\bibnamefont {Read}},\
  }\href {\doibase DOI: 10.1016/0550-3213(91)90407-O} {\bibfield  {journal}
  {\bibinfo  {journal} {Nuclear Physics B}\ }\textbf {\bibinfo {volume}
  {360}},\ \bibinfo {pages} {362 } (\bibinfo {year} {1991})}\BibitemShut
  {NoStop}%
\bibitem [{\citenamefont {Carusotto}\ and\ \citenamefont
  {Mueller}(2004)}]{Carusotto2004a}%
  \BibitemOpen
  \bibfield  {author} {\bibinfo {author} {\bibfnamefont {I.}~\bibnamefont
  {Carusotto}}\ and\ \bibinfo {author} {\bibfnamefont {E.~J.}\ \bibnamefont
  {Mueller}},\ }\href {http://stacks.iop.org/0953-4075/37/i=7/a=058} {\bibfield
   {journal} {\bibinfo  {journal} {Journal of Physics B: Atomic, Molecular and
  Optical Physics}\ }\textbf {\bibinfo {volume} {37}},\ \bibinfo {pages} {S115}
  (\bibinfo {year} {2004})}\BibitemShut {NoStop}%
\bibitem [{\citenamefont {Higbie}\ \emph {et~al.}(2005)\citenamefont {Higbie},
  \citenamefont {Sadler}, \citenamefont {Inouye}, \citenamefont {Chikkatur},
  \citenamefont {Leslie}, \citenamefont {Moore}, \citenamefont {Savalli},\ and\
  \citenamefont {Stamper-Kurn}}]{Higbie2005a}%
  \BibitemOpen
  \bibfield  {author} {\bibinfo {author} {\bibfnamefont {J.~M.}\ \bibnamefont
  {Higbie}}, \bibinfo {author} {\bibfnamefont {L.~E.}\ \bibnamefont {Sadler}},
  \bibinfo {author} {\bibfnamefont {S.}~\bibnamefont {Inouye}}, \bibinfo
  {author} {\bibfnamefont {A.~P.}\ \bibnamefont {Chikkatur}}, \bibinfo {author}
  {\bibfnamefont {S.~R.}\ \bibnamefont {Leslie}}, \bibinfo {author}
  {\bibfnamefont {K.~L.}\ \bibnamefont {Moore}}, \bibinfo {author}
  {\bibfnamefont {V.}~\bibnamefont {Savalli}}, \ and\ \bibinfo {author}
  {\bibfnamefont {D.~M.}\ \bibnamefont {Stamper-Kurn}},\ }\href {\doibase
  10.1103/PhysRevLett.95.050401} {\bibfield  {journal} {\bibinfo  {journal}
  {Phys. Rev. Lett.}\ }\textbf {\bibinfo {volume} {95}},\ \bibinfo {pages}
  {050401} (\bibinfo {year} {2005})}\BibitemShut {NoStop}%
\bibitem [{\citenamefont {Carusotto}(2006)}]{Carusotto2006a}%
  \BibitemOpen
  \bibfield  {author} {\bibinfo {author} {\bibfnamefont {I.}~\bibnamefont
  {Carusotto}},\ }\href {http://stacks.iop.org/0953-4075/39/i=10/a=S20}
  {\bibfield  {journal} {\bibinfo  {journal} {Journal of Physics B: Atomic,
  Molecular and Optical Physics}\ }\textbf {\bibinfo {volume} {39}},\ \bibinfo
  {pages} {S211} (\bibinfo {year} {2006})}\BibitemShut {NoStop}%
\bibitem [{\citenamefont {Eckert}\ \emph {et~al.}(2007)\citenamefont {Eckert},
  \citenamefont {Zawitkowski}, \citenamefont {Sanpera}, \citenamefont
  {Lewenstein},\ and\ \citenamefont {Polzik}}]{Eckert2007a}%
  \BibitemOpen
  \bibfield  {author} {\bibinfo {author} {\bibfnamefont {K.}~\bibnamefont
  {Eckert}}, \bibinfo {author} {\bibfnamefont {L.}~\bibnamefont {Zawitkowski}},
  \bibinfo {author} {\bibfnamefont {A.}~\bibnamefont {Sanpera}}, \bibinfo
  {author} {\bibfnamefont {M.}~\bibnamefont {Lewenstein}}, \ and\ \bibinfo
  {author} {\bibfnamefont {E.~S.}\ \bibnamefont {Polzik}},\ }\href {\doibase
  10.1103/PhysRevLett.98.100404} {\bibfield  {journal} {\bibinfo  {journal}
  {Phys. Rev. Lett.}\ }\textbf {\bibinfo {volume} {98}},\ \bibinfo {pages}
  {100404} (\bibinfo {year} {2007})}\BibitemShut {NoStop}%
\bibitem [{\citenamefont {de~Vega}\ \emph {et~al.}(2008)\citenamefont
  {de~Vega}, \citenamefont {Cirac},\ and\ \citenamefont {Porras}}]{Vega2008a}%
  \BibitemOpen
  \bibfield  {author} {\bibinfo {author} {\bibfnamefont {I.}~\bibnamefont
  {de~Vega}}, \bibinfo {author} {\bibfnamefont {J.~I.}\ \bibnamefont {Cirac}},
  \ and\ \bibinfo {author} {\bibfnamefont {D.}~\bibnamefont {Porras}},\ }\href
  {\doibase 10.1103/PhysRevA.77.051804} {\bibfield  {journal} {\bibinfo
  {journal} {Phys. Rev. A}\ }\textbf {\bibinfo {volume} {77}},\ \bibinfo
  {pages} {051804} (\bibinfo {year} {2008})}\BibitemShut {NoStop}%
\bibitem [{\citenamefont {Eckert}\ \emph {et~al.}(2008)\citenamefont {Eckert},
  \citenamefont {Romero-Isart}, \citenamefont {Rodr\'iguez}, \citenamefont
  {Lewenstein}, \citenamefont {Polzik},\ and\ \citenamefont
  {Sanpera}}]{Eckert2008a}%
  \BibitemOpen
  \bibfield  {author} {\bibinfo {author} {\bibfnamefont {K.}~\bibnamefont
  {Eckert}}, \bibinfo {author} {\bibfnamefont {O.}~\bibnamefont
  {Romero-Isart}}, \bibinfo {author} {\bibfnamefont {M.}~\bibnamefont
  {Rodr\'iguez}}, \bibinfo {author} {\bibfnamefont {M.}~\bibnamefont
  {Lewenstein}}, \bibinfo {author} {\bibfnamefont {E.~S.}\ \bibnamefont
  {Polzik}}, \ and\ \bibinfo {author} {\bibfnamefont {A.}~\bibnamefont
  {Sanpera}},\ }\href@noop {} {\bibfield  {journal} {\bibinfo  {journal} {Nat.
  Phys.}\ }\textbf {\bibinfo {volume} {4}},\ \bibinfo {pages} {50} (\bibinfo
  {year} {2008})}\BibitemShut {NoStop}%
\bibitem [{\citenamefont {Zhang}\ \emph {et~al.}(2009)\citenamefont {Zhang},
  \citenamefont {Cui}, \citenamefont {Jing}, \citenamefont {Zhou},\ and\
  \citenamefont {Liu}}]{Zhang2009a}%
  \BibitemOpen
  \bibfield  {author} {\bibinfo {author} {\bibfnamefont {J.~M.}\ \bibnamefont
  {Zhang}}, \bibinfo {author} {\bibfnamefont {S.}~\bibnamefont {Cui}}, \bibinfo
  {author} {\bibfnamefont {H.}~\bibnamefont {Jing}}, \bibinfo {author}
  {\bibfnamefont {D.~L.}\ \bibnamefont {Zhou}}, \ and\ \bibinfo {author}
  {\bibfnamefont {W.~M.}\ \bibnamefont {Liu}},\ }\href {\doibase
  10.1103/PhysRevA.80.043623} {\bibfield  {journal} {\bibinfo  {journal} {Phys.
  Rev. A}\ }\textbf {\bibinfo {volume} {80}},\ \bibinfo {eid} {043623}
  (\bibinfo {year} {2009})}\BibitemShut {NoStop}%
\bibitem [{\citenamefont {Roscilde}\ \emph {et~al.}(2009)\citenamefont
  {Roscilde}, \citenamefont {Rodr\'iguez}, \citenamefont {Eckert},
  \citenamefont {Romero-Isart}, \citenamefont {Lewenstein}, \citenamefont
  {Polzik},\ and\ \citenamefont {Sanpera}}]{Roscilde2009a}%
  \BibitemOpen
  \bibfield  {author} {\bibinfo {author} {\bibfnamefont {T.}~\bibnamefont
  {Roscilde}}, \bibinfo {author} {\bibfnamefont {M.}~\bibnamefont
  {Rodr\'iguez}}, \bibinfo {author} {\bibfnamefont {K.}~\bibnamefont {Eckert}},
  \bibinfo {author} {\bibfnamefont {O.}~\bibnamefont {Romero-Isart}}, \bibinfo
  {author} {\bibfnamefont {M.}~\bibnamefont {Lewenstein}}, \bibinfo {author}
  {\bibfnamefont {E.}~\bibnamefont {Polzik}}, \ and\ \bibinfo {author}
  {\bibfnamefont {A.}~\bibnamefont {Sanpera}},\ }\href
  {http://stacks.iop.org/1367-2630/11/i=5/a=055041} {\bibfield  {journal}
  {\bibinfo  {journal} {New J. Phys.}\ }\textbf {\bibinfo {volume} {11}},\
  \bibinfo {pages} {055041} (\bibinfo {year} {2009})}\BibitemShut {NoStop}%
\bibitem [{\citenamefont {Javanainen}\ and\ \citenamefont
  {Ruostekoski}(2003)}]{Javanainen2003}%
  \BibitemOpen
  \bibfield  {author} {\bibinfo {author} {\bibfnamefont {J.}~\bibnamefont
  {Javanainen}}\ and\ \bibinfo {author} {\bibfnamefont {J.}~\bibnamefont
  {Ruostekoski}},\ }\href {\doibase 10.1103/PhysRevLett.91.150404} {\bibfield
  {journal} {\bibinfo  {journal} {Phys. Rev. Lett.}\ }\textbf {\bibinfo
  {volume} {91}},\ \bibinfo {pages} {150404} (\bibinfo {year}
  {2003})}\BibitemShut {NoStop}%
\bibitem [{\citenamefont {Ruostekoski}\ \emph {et~al.}(2008)\citenamefont
  {Ruostekoski}, \citenamefont {Javanainen},\ and\ \citenamefont
  {Dunne}}]{Ruostekoski2008a}%
  \BibitemOpen
  \bibfield  {author} {\bibinfo {author} {\bibfnamefont {J.}~\bibnamefont
  {Ruostekoski}}, \bibinfo {author} {\bibfnamefont {J.}~\bibnamefont
  {Javanainen}}, \ and\ \bibinfo {author} {\bibfnamefont {G.~V.}\ \bibnamefont
  {Dunne}},\ }\href {\doibase 10.1103/PhysRevA.77.013603} {\bibfield  {journal}
  {\bibinfo  {journal} {Phys. Rev. A}\ }\textbf {\bibinfo {volume} {77}},\
  \bibinfo {pages} {013603} (\bibinfo {year} {2008})}\BibitemShut {NoStop}%
\bibitem [{\citenamefont {Douglas}\ and\ \citenamefont
  {Burnett}(2010)}]{Douglas2010a}%
  \BibitemOpen
  \bibfield  {author} {\bibinfo {author} {\bibfnamefont {J.~S.}\ \bibnamefont
  {Douglas}}\ and\ \bibinfo {author} {\bibfnamefont {K.}~\bibnamefont
  {Burnett}},\ }\href {\doibase 10.1103/PhysRevA.82.033434} {\bibfield
  {journal} {\bibinfo  {journal} {Phys. Rev. A}\ }\textbf {\bibinfo {volume}
  {82}},\ \bibinfo {pages} {033434} (\bibinfo {year} {2010})}\BibitemShut
  {NoStop}%
\bibitem [{\citenamefont {Weitenberg}\ \emph {et~al.}(2011)\citenamefont
  {Weitenberg}, \citenamefont {Schau\ss{}}, \citenamefont {Fukuhara},
  \citenamefont {Cheneau}, \citenamefont {Endres}, \citenamefont {Bloch},\ and\
  \citenamefont {Kuhr}}]{Weitenberg2011a}%
  \BibitemOpen
  \bibfield  {author} {\bibinfo {author} {\bibfnamefont {C.}~\bibnamefont
  {Weitenberg}}, \bibinfo {author} {\bibfnamefont {P.}~\bibnamefont
  {Schau\ss{}}}, \bibinfo {author} {\bibfnamefont {T.}~\bibnamefont
  {Fukuhara}}, \bibinfo {author} {\bibfnamefont {M.}~\bibnamefont {Cheneau}},
  \bibinfo {author} {\bibfnamefont {M.}~\bibnamefont {Endres}}, \bibinfo
  {author} {\bibfnamefont {I.}~\bibnamefont {Bloch}}, \ and\ \bibinfo {author}
  {\bibfnamefont {S.}~\bibnamefont {Kuhr}},\ }\href {\doibase
  10.1103/PhysRevLett.106.215301} {\bibfield  {journal} {\bibinfo  {journal}
  {Phys. Rev. Lett.}\ }\textbf {\bibinfo {volume} {106}},\ \bibinfo {pages}
  {215301} (\bibinfo {year} {2011})}\BibitemShut {NoStop}%
\bibitem [{\citenamefont {Hammerer}\ \emph {et~al.}(2010)\citenamefont
  {Hammerer}, \citenamefont {S\o{}rensen},\ and\ \citenamefont
  {Polzik}}]{Hammerer2010a}%
  \BibitemOpen
  \bibfield  {author} {\bibinfo {author} {\bibfnamefont {K.}~\bibnamefont
  {Hammerer}}, \bibinfo {author} {\bibfnamefont {A.~S.}\ \bibnamefont
  {S\o{}rensen}}, \ and\ \bibinfo {author} {\bibfnamefont {E.~S.}\ \bibnamefont
  {Polzik}},\ }\href {\doibase 10.1103/RevModPhys.82.1041} {\bibfield
  {journal} {\bibinfo  {journal} {Rev. Mod. Phys.}\ }\textbf {\bibinfo {volume}
  {82}},\ \bibinfo {pages} {1041} (\bibinfo {year} {2010})}\BibitemShut
  {NoStop}%
\bibitem [{\citenamefont {Deutsch}\ and\ \citenamefont
  {Jessen}(1998)}]{Deutsch1998a}%
  \BibitemOpen
  \bibfield  {author} {\bibinfo {author} {\bibfnamefont {I.~H.}\ \bibnamefont
  {Deutsch}}\ and\ \bibinfo {author} {\bibfnamefont {P.~S.}\ \bibnamefont
  {Jessen}},\ }\href {\doibase 10.1103/PhysRevA.57.1972} {\bibfield  {journal}
  {\bibinfo  {journal} {Phys. Rev. A}\ }\textbf {\bibinfo {volume} {57}},\
  \bibinfo {pages} {1972} (\bibinfo {year} {1998})}\BibitemShut {NoStop}%
\bibitem [{\citenamefont {Douglas}\ and\ \citenamefont
  {Burnett}(2011)}]{Douglas2011a}%
  \BibitemOpen
  \bibfield  {author} {\bibinfo {author} {\bibfnamefont {J.~S.}\ \bibnamefont
  {Douglas}}\ and\ \bibinfo {author} {\bibfnamefont {K.}~\bibnamefont
  {Burnett}},\ }\href {\doibase 10.1103/PhysRevA.84.033637} {\bibfield
  {journal} {\bibinfo  {journal} {Phys. Rev. A}\ }\textbf {\bibinfo {volume}
  {84}},\ \bibinfo {pages} {033637} (\bibinfo {year} {2011})}\BibitemShut
  {NoStop}%
\bibitem [{\citenamefont {{Douglas}}\ and\ \citenamefont
  {{Burnett}}(2011)}]{Douglas2011b}%
  \BibitemOpen
  \bibfield  {author} {\bibinfo {author} {\bibfnamefont {J.~S.}\ \bibnamefont
  {{Douglas}}}\ and\ \bibinfo {author} {\bibfnamefont {K.}~\bibnamefont
  {{Burnett}}},\ }\href@noop {} {\bibfield  {journal} {\bibinfo  {journal}
  {ArXiv e-prints}\ } (\bibinfo {year} {2011})},\ \Eprint
  {http://arxiv.org/abs/1109.0041} {arXiv:1109.0041 [quant-ph]} \BibitemShut
  {NoStop}%
\bibitem [{\citenamefont {Newman}\ and\ \citenamefont
  {Barkema}(1999)}]{Newman1999a}%
  \BibitemOpen
  \bibfield  {author} {\bibinfo {author} {\bibfnamefont {M.~E.~J.}\
  \bibnamefont {Newman}}\ and\ \bibinfo {author} {\bibfnamefont {G.~T.}\
  \bibnamefont {Barkema}},\ }\href@noop {} {\emph {\bibinfo {title} {Monte
  Carlo methods in statistical physics}}}\ (\bibinfo  {publisher} {Oxford
  University Press},\ \bibinfo {year} {1999})\BibitemShut {NoStop}%
\end{thebibliography}
%
%
\end{document}